\documentclass[preprint,superscriptaddress,secnumarabic,amssymb, nobibnotes, aps, pre]{revtex4-2}
\usepackage{makecell}
\usepackage{bm}
\usepackage{mathrsfs}
\usepackage{amsmath}
\usepackage{tikz}
\usetikzlibrary{automata,patterns,positioning}
\usepackage{graphicx}
\usepackage{pgfplots}
\usepgfplotslibrary{external}
\pgfplotsset{compat=newest}
\pgfplotsset{scaled y ticks=false}
\usetikzlibrary{plotmarks}
\usetikzlibrary{arrows.meta}
\usetikzlibrary{3d}
\usepgfplotslibrary{patchplots}
\usepackage{grffile}
\pgfplotsset{plot coordinates/math parser=false}
\newlength\figureheight
\newlength\figurewidth
\usepgfplotslibrary{colormaps}
\usepackage[siunitx]{circuitikz}

\usepackage{tikz}
\usepackage{tikzsymbols}
\usepackage{titletoc}
\usetikzlibrary{shapes.geometric, arrows, decorations.pathreplacing, patterns, decorations.markings}

\usetikzlibrary{fpu}
\usepackage{xintexpr}
\usepackage{pifont}
\usetikzlibrary{intersections}
\pgfdeclarelayer{background}
\pgfsetlayers{background,main}

\graphicspath{ {Figures/} }

\newcommand\Fig{Figure} 
\newcommand\fig{Fig.}  
\newcommand\figs{Figs.}  

\newcommand{\vect}[1]{\boldsymbol{#1}}
\newcommand{\pde}[2]{\frac{\partial #1}{\partial #2}}

\newcommand\bnabla{\vect{\nabla}}
\newcommand\bcdot{\vect{\cdot}}

\newcommand\ii{\mathrm{i}}
\newcommand\xhp{\hat{\xi}_\perp}

\def\mcy{black}

\newcommand\bnp{\bnabla_\perp}
\newcommand\pp{p_\perp}
\newcommand\Pp{P_\perp}
\newcommand\uvp{\vect{u}_\perp}
\newcommand\Uvp{\vect{U}_\perp}
\newcommand\up{u_\perp}
\newcommand\Up{U_\perp}
\newcommand\vp{v_\perp}
\newcommand\p{\partial}

\newcommand\xc{\check{x}}
\newcommand\yc{\check{y}}
\newcommand\zc{\check{z}}
\newcommand\tc{\check{t}}

\newcommand\upp{u_\perp'}
\newcommand\vpp{v_\perp'}
\newcommand\ppp{p_\perp'}
\newcommand\uvpp{\vect{u}_\perp'}
\newcommand\uhp{\hat{u}_\perp}
\newcommand\vhp{\hat{v}_\perp}

\newcommand\uvph{\hat{\vect{u}}_\perp}

\newcommand\Rey{\mathrm{Re}}  
\newcommand\Sr{\mathrm{Sr}}
\newcommand\Wo{\mathrm{Wo}}
\newcommand\Rez{\mathrm{Real}}

\newcommand\Rm{R_\mathrm{m}}

\newcommand\ReyG{\mathrm{Re}_\Gamma}
\newcommand\ReyCrit{\mathrm{Re}_\mathrm{c}}
\newcommand\ReyCritPS{\mathrm{Re}_\mathrm{c,ps}}
\newcommand\ReyCritST{\mathrm{Re}_\mathrm{c,st}}
\newcommand\rrs{r_\mathrm{s}}
\newcommand\rrc{r_\mathrm{c}}

\newcommand\rrpc{r_\mathrm{c,ps}}

\newcommand\Nc{N_\mathrm{c}}
\newcommand\Np{N_\mathrm{p}}
\newcommand\tzero{t_0}
\newcommand\Gmax{G_\mathrm{max}}
\newcommand\tzeroOpt{t_{0,\mathrm{opt}}}
\newcommand\tzeroB{t_{0,\mathrm{B16}}}
\newcommand\alphaOpt{\alpha_\mathrm{opt}}
\newcommand\tauOpt{\tau_\mathrm{opt}}
\newcommand\amax{\alpha_\mathrm{max}}

\newcommand\IC{I_\mathrm{max}}
\newcommand\gst{G_\mathrm{s}}
\newcommand\astp{\alpha_\mathrm{s}}
\newcommand\tst{\tau_\mathrm{s}}
\newcommand\tzn{t_{0,\mathrm{s}}}
\newcommand\GmaxST{G_\mathrm{max,st}}
\newcommand\alphaST{\alpha_\mathrm{opt,st}}
\newcommand\tauST{\tau_\mathrm{opt,st}}

\newcommand\Ev{E_\mathrm{v}}
\newcommand\Ezero{E_0}

\begin{document}

\title{Pulsatility delays the transition to sustained turbulence in quasi-two-dimensional shear flows}%

\author{Christopher J. Camobreco}%
\email{christopher.camobreco@unimelb.edu.au}
\affiliation{Department of Mechanical and Aerospace Engineering, Monash University, Victoria 3800, Australia}
\affiliation{Department of Mechanical Engineering, University of Melbourne, Victoria 3010, Australia}
\author{Alban Poth{\'e}rat}%
\email{alban.potherat@coventry.ac.uk}
\affiliation{Fluid and Complex Systems Research Centre, Coventry University, Coventry CV15FB, United Kingdom}
\author{Gregory J. Sheard}%
\email{greg.sheard@monash.edu}
\affiliation{Department of Mechanical and Aerospace Engineering, Monash University, Victoria 3800, Australia}
\date{\today}%

\begin{abstract}
This work investigates efficient routes to turbulence in quasi-two-dimensional shear flows.
Two-dimensional disturbances require high Reynolds numbers to incite transition from a steady base flow, as transient growth is modest. With the addition of an oscillatory base flow component, this work shows that the transient growth experienced by two-dimensional initial perturbations is often well above that provided by the steady component. However, as has been shown for three-dimensional flows [Pier \& Schmid J.~Fluid Mech.~\textbf{926}, A11 (2021)], the transient growth is almost entirely composed of modal intracyclic growth, rather than a transient mechanism which takes advantage of non-normality. This lack of transient growth, relative to the severe decay induced by the favorable pressure gradient during the acceleration phase of the oscillatory base flow, only ever delays the transition to sustained turbulence. Thus, a non-oscillatory driving force remains the most efficient strategy for sustained turbulence in quasi-two-dimensional shear flows. The only benefit provided by pulsatility is that the amplitude of the initial condition required to trigger intermittent turbulence is orders of magnitude smaller.
\end{abstract}

\maketitle

\section{Introduction} \label{sec:intro}

Turbulent flows enhance momentum transport and thereby improve scalar mixing \citep{Falkovich2018fluid}. Turbulent mixing is desirable, provided that the flow rates necessary to induce turbulence are not prohibitive, in the rebreeder-coolant conduits of magnetic confinement fusion reactors \citep{Barleon2000heat}. However, at fusion relevant conditions, the flow is anticipated to be quasi-two-dimensional (Q2D) in the idealized configuration of a straight rectangular pipe, due to the interaction between the pervading transverse magnetic field and electrically conducting rebreeder fluid \citep{Smolentsev2008characterization}. Variations in flow quantities are minor along magnetic field lines, except in asymptotically thin exponential boundary layers \citep{Sommeria1982why, Potherat2000effective, Muller2001magnetofluiddynamics}, in the limit of high magnetic fields. Numerical simulations approaching fusion relevant field strengths (Q2D regimes) indicate that the three-dimensional lift-up and oblique-wave mechanisms generate negligible linear transient growth \citep{Cassels2019from3D}. Linear growth is almost solely generated by the two-dimensional Orr mechanism, in excellent agreement with predictions from a Q2D model \citep{Potherat2007quasi}. Thus, flows in Q2D regimes cannot rely on the large linear growth often provided by the lift-up mechanism \citep{Butler1992optimal,Trefethen1993hydrodynamic,Schmid2001stability}, with implications on inciting and sustaining turbulence \citep{Lozano2020cause}. In Q2D duct flows, transitions to sustained turbulence have only been observed at large, weakly subcritical, Reynolds numbers (within 20\% of the critical Reynolds number obtained by linear stability analysis) \citep{Camobreco2021weakly}.

As a steady base flow may only sustain turbulence at Reynolds numbers too large for practical purposes \citep{Smolentsev2010considerations,Smolentsev2010mhd}, the application of an oscillatory driving force with an underlying steady pressure gradient has been investigated \citep{Camobreco2021stability}. The pulsatile (steady $+$ oscillatory) Q2D base flows yielded over an order of magnitude reduction in the critical Reynolds numbers, relative to their steady counterparts, with greater percentage destabilizations than hydrodynamic studies \citep{Thomas2011linear,Pier2017linear}. However, these Q2D pulsatile flows did not sustain turbulence, even at critical and weakly supercritical Reynolds numbers (relative to the pulsatile base flow).

While an oscillatory driving force may aid the transition to turbulence, it can equally present difficulties to turbulence sustainment. Perturbation growth predominantly occurs during the deceleration phase of the base flow \citep{Biau2016transient,Pier2017linear}, with laminar-turbulent transitions consistently observed near the end of the deceleration phase \citep{Tuzi2008intermittent,Ozdemir2014direct,Ebadi2019mean,Xu2020nonlinear}. However, during the acceleration phase, the flow may relaminarize (at least partially), leading to temporally intermittent turbulence which is reinvigorated each cycle by intracyclic perturbation growth. Fourier spectra indicating fluctuation energies with possible $-5/3$ power law dependence on wave number, as for fully developed Q2D or 3D turbulence \citep{Pope2000turbulent}, have only been observed at certain phases through the cycle \citep{Tuzi2008intermittent,Ozdemir2014direct}. Overall, in both simple and complex geometries, turbulent fluctuations have typically only been observed over the large scale perturbation energy variations due to intracyclic growth and decay \citep{Ozdemir2014direct,Sherwin2005three}. Thus, sustained turbulence has rarely been observed in pulsatile flows, even though pulsatility continues to provide significant perturbation growth potential after the turbulent transition. The first aim of this work is thereby to extend the investigation of Ref.~\citep{Camobreco2021stability}, to establish whether the addition of an oscillatory component to the base flow ever yields sustained turbulence at a Reynolds number lower than for the steady counterpart. To do so, initial conditions optimizing transient growth will be investigated, rather than just white noise initial conditions.

Pulsatility has also been shown to have a significant effect on modal growth mechanisms while simultaneously having a negligible effect on non-normality, and thereby non-modal transient growth, in hydrodynamic flows \citep{Kern2021transient,Kern2022subharmonic,Pier2021optimal}. Even at intermediate frequencies (those of the most interest), Ref.~\citep{Pier2021optimal} conclude that modal transient growth provides the predominant contribution even to optimal non-modal transient growth, which occurs over approximately half a pulsation cycle.  With optimized perturbations, the maximum transient growth depends exponentially on Reynolds number \citep{Biau2016transient,Xu2021non}, as does the intracylcic growth of modal instabilities \citep{Camobreco2021stability,Pier2017linear}, with similar trends observed in more complex geometries \citep{Blackburn2008pulsatile}. The total growth is then the linear combination of the transient growth from the steady base flow component, and modal intracyclic growth from the oscillating component. By comparison, at high and low frequencies, transient growth is predominantly due to the steady component, the oscillations providing little assistance in generating net growth. While other non-periodic non-modal transient growth measurements have been performed \citep{Biau2016transient,Tsigklifis2017asymptotic,Xu2021non}, only Ref.~\citep{Tsigklifis2017asymptotic} makes direct comparisons between modal transient growth, and non-periodic non-modal transient growth, and not in the frequency range of interest. Using a different framework, Ref.~\citep{Kern2021transient} optimized for the perturbation with growth rate maximized at each instant in time. This allowed measurement of the non-modal growth potential each cycle, rather than measuring non-periodic non-modal transient growth. Although only a small parameter range was investigated, a similar conclusion was reached, that non-modal growth is not enhanced by pulsatility.

While optimized non-modal perturbations in pulsatile flows are expected to behave similarly to their steady counterparts, one key difference has been previously observed. For an isolated oscillatory boundary layer, the optimal out-of-plane wave number was found to be zero for all Reynolds numbers, indicating that transient growth arises solely due to the Orr mechanism \citep{Biau2016transient}.  At low pulsation amplitudes, toward the purely steady limit, transient growth of 3D optimals is maximized by the lift-up mechanism \citep{Pier2021optimal}, based on the steady base flow component. However, at larger frequencies and amplitudes, toward the purely oscillatory limit, the Orr mechanism (or an axisymmetric Orr-like mechanism for pipe flows) maximizes transient growth of now 2D (or axisymmetric) optimals \citep{Pier2021optimal,Xu2021non}. The optimal growth of these 2D instabilities is believed to be due to maximized intracyclic (modal) growth \citep{Pier2021optimal}. This hypothesis will be further examined in this work. As the optimal perturbations are naturally 2D at these larger frequencies and amplitudes, a study of Q2D flows in this parameter range does not occlude the relevant linear growth mechanisms. 
Thus, the second aim of this work is to identify the overall importance of non-modal transient growth, relative to the inherent modal transient growth dynamics of pulsatile base flows. Furthermore a systematic investigation of the transient growth dynamics over a wider range of pulsation frequencies and amplitudes, as well as larger Reynolds numbers and friction parameters (magnetic field strengths), will be performed.

This work proceeds as follows. Sec.~\ref{sec:probl} defines the problem, and discusses the applicability of the Q2D model for pulsatile duct flows. Sec.~\ref{sec:ltgro} covers the non-modal linear transient growth computations, over a wide range of pulsation frequencies, at large and intermediate amplitude ratios, and compares modal intracyclic growth to non-modal transient growth. Fully nonlinear direct numerical simulations are then analysed. DNS targeting non-modal instabilities are presented in Sec.~\ref{sec:nlinm}, and their ability to sustain turbulence compared to the best-case steady counterpart is assessed. In addition, DNS targeting modal instabilities are revisited, to highlight the complications introduced at larger pulsation amplitudes. Conclusions follow in Sec.~\ref{sec:concl}. Note henceforth that ``intracyclic growth'' (intracyclic normal-mode growth in Ref.~\citep{Pier2021optimal}) refers to modal transient growth, while ``transient growth'' refers to conventional non-modal non-periodic growth \citep{Schmid2001stability}.

\section{Problem Setup} \label{sec:probl}

\subsection{Setup and Q2D model validity}

\begin{figure}[h!]
\centering
\includegraphics[]{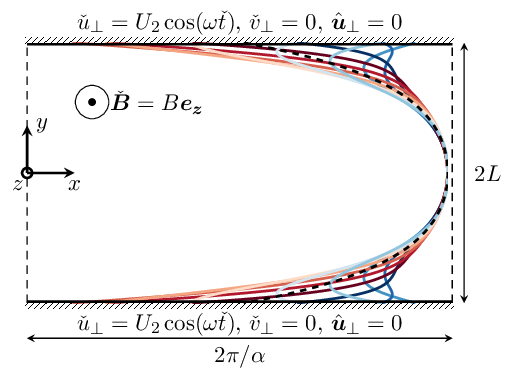}
\vspace{-7mm}
\caption{A schematic representation of the system. Solid lines denote oscillating, impermeable, no-slip walls. Short dashed lines indicate the streamwise extent of the periodic domain, defined by streamwise wave number $\alpha$. An example base flow at $H=10$, $\Gamma=1.19$, $\Sr = 5.4\times10^{-3}$ and $\Rey = 1.5\times10^4$ is overlaid; the dashed line denotes the steady base flow component and the 11 colored lines the normalized total pulsatile base flow $\Up(t,y)/\gamma_1$ equispaced over $0\leq t\leq 2\pi$.}
\label{fig:prob_set}
\end{figure}

This work investigates the magnetohydrodynamic (MHD) flow of an electrically conducting incompressible Newtonian fluid, through a streamwise invariant duct of rectangular cross-section, subjected to a uniform transverse magnetic field $B\vect{e_z}$, see \fig~\ref{fig:prob_set}. The fluid has density $\rho$, kinematic viscosity $\nu$ and electrical conductivity $\sigma$. The wall-normal duct height is $2L$ ($\yc$ direction) and transverse duct width $a$ ($\zc$ direction). The $\xc$ direction is periodic, with streamwise wave number $\alpha$. All walls are no-slip, impermeable and electrically insulating. A constant pressure gradient drives a steady base flow component, with maximum undisturbed dimensional velocity $U_1$. Synchronous oscillation of both lateral walls at velocity $U_2\cos(\omega \tc)$, with frequency $\omega$ and maximum dimensional velocity $U_2$, drives an oscillatory base flow component. The pulsatile flow, the sum of the steady and oscillatory components, has a maximum velocity over the cycle of $U_0$. Assuming the flow is quasi-two-dimensional, the incompressible MHD equations in the quasi-static limit \citep{Roberts1967introduction}, averaged along $\vect{e_z}$, yield a 2D set of equations \citep{Sommeria1982why}. With length, velocity, time and pressure non-dimensionalized by $L$, $U_0$, $1/\omega$ and $\rho U_0^2$, respectively, this set of equations becomes \citep{Potherat2007quasi,Sommeria1982why}:
\begin{equation}\label{eq:Q2Dc}
\bnp \bcdot \uvp = 0,
\end{equation}
\begin{equation} \label{eq:Q2Dm}
\Sr\pde{\uvp}{t} = - (\uvp \bcdot \bnp) \uvp - \bnp \pp + \frac{1}{\Rey}\bnp^2 \uvp - \frac{H}{\Rey} \uvp,
\end{equation}
where $\uvp=(\up$, $\vp)$, $\bnp = (\p_x$, $\p_y)$, the Reynolds number $\Rey = U_0 L/\nu$, the Strouhal number $\Sr = \omega L /U_0$ and the Hartmann friction parameter $H=2B(L^2/a)(\sigma/\rho\nu)^{1/2}$. \textcolor{\mcy}{Respectively, these parameters represent the dimensionless ratios of steady inertial to viscous forces ($\Rey$), transient inertial to steady inertial forces ($\Sr$) and (aspect-ratio adjusted) electromagnetic to viscous forces ($H$).} Although the Womersly number $\Wo^2 = \Sr\Rey$ can be used instead of $\Sr$ as a dimensionless frequency, such a choice would impact the ability to vary $\Rey$ at constant pulsation frequency (constant $\Sr$). For consistency with Refs.~\citep{Thomas2011linear,Camobreco2021stability}, the non-dimensional oscillatory wall boundary conditions are $\uvp(y\pm1) = (\gamma_2\cos(t)/\gamma_1$, $0)$ and the constant driving pressure gradient $\gamma_1 \p_x \pp |_{y=0}$, where $\gamma_1=\Gamma/(1+\Gamma)$, $\gamma_2=1/(1+\Gamma)$ and the amplitude ratio $\Gamma = U_1/U_2$, \textcolor{\mcy}{where $\Gamma$ is the ratio of the maximum velocity of the undisturbed (steady) flow $U_1$ relative to that of the oscillating flow $U_2$. Thus,} $\Gamma=0$ represents a flow purely driven by oscillating walls (no pressure gradient; $\gamma_1=0$, $\gamma_2=1$) and $\Gamma \rightarrow \infty$ a pressure driven flow (no wall motion; $\gamma_1\rightarrow1$, $\gamma_2\rightarrow0$). Large amplitude ratios $\Gamma$ \textcolor{\mcy}{thereby} represent small amplitudes of the pulsatile component, unlike recent comparative studies which take the inverse definition \citep{Pier2017linear,Pier2021optimal,Kern2021transient,Kern2022subharmonic}. As Eq.~(\ref{eq:Q2Dm}) satisfies extended Galilean invariance \citep{Pope2000turbulent}, the use of oscillatory wall motion to drive the flow is exactly equivalent to using an oscillatory pressure gradient \citep{Camobreco2021stability}, had the latter been preferred.

Numerous works have derived and validated Q2D models for 3D MHD duct flows \citep{Potherat2000effective,Cassels2019from3D,Kanaris2013numerical, Muck2000flows,Dousset2008numerical,Potherat2007quasi,Sommeria1982why}. However, an oscillatory driving force introduces additional constraints on the validity of the Q2D model, as discussed in detail in Ref.~\citep{Camobreco2021stability}, and recapped here.  First, pulsatility introduces a shielding constraint of $\Sr \ll \Rm^{-1}$, where the magnetic Reynolds number $\Rm=\mu_0\sigma U_1 L$ and where $\mu_0$ is the permeability of free space. This ensures that the magnetic field can diffuse throughout the entire domain, and translates to $\Sr \ll 100$ for the liquid metals of interest \citep{Knaepen2004magnetohydrodynamic,Potherat2015decay}. Second, to ensure the magnetic field can be treated as steady, the Alfv{\'e}n timescale must be much smaller than the pulsation timescale. This requires $\Sr(U_0/U_1)<1$, or $\Sr/\gamma_1<1$, and translates to $\Sr<1/2$ for the $\Gamma \geq 1$ of interest \citep{Camobreco2021stability}. Note that this also assumes that quasi-two-dimensionality is valid for the equivalent steady flow characterised by $U_1$. Third, to ensure diffusion of momentum along magnetic field lines occurs more rapidly than any in-plane momentum transfers (i.e.~the quasi-two-dimensionalization of the flow) requires $\Sr(U_0/U_1)<1$, or again that $\Sr<1/2$.

\subsection{Perturbations and the base flow}

Perturbations $(\uvpp$, $\ppp)$ are defined as, 
\begin{equation}\label{eq:def_pert}
\uvpp = \uvp - \Uvp, \,\,\, \ppp = \pp - \Pp,
\end{equation}
to focus on the deviation between the instantaneous flow $(\uvp$, $\pp)$ and the laminar base flow $(\Uvp$, $\Pp)$. Thus, perturbations can be of any magnitude, 
and will be used to identify turbulent flows. Only perturbations with magnitudes $\ll 1$ permit linearization.

The fully developed, parallel, pulsatile base flow $\Uvp=\Up\vect{e_x} = \Up(t+2\pi,y)\vect{e_x}$ satisfying Eqs.~(\ref{eq:Q2Dc}) and (\ref{eq:Q2Dm}) is
\begin{equation}\label{eq:def_base}
\Up = \gamma_1\frac{\cosh(H^{1/2})}{\cosh(H^{1/2})-1}\left(1-\frac{\cosh(H^{1/2}y)}{\cosh(H^{1/2})}\right) + \gamma_2\Rez\!\left(\frac{\cosh[(r+s\mathrm{i})y]}{\cosh(r+s\mathrm{i})}e^{\mathrm{i}t}\right),
\end{equation}
where $\mathrm{i}=(-1)^{1/2}$. The base flow is driven by a constant streamwise pressure gradient $\p_x\Pp = -\gamma_1(H/\Rey)\cosh(H^{1/2})/[\cosh(H^{1/2})-1]$ and satisfies lateral wall oscillations of $\Up(y \pm 1) = \gamma_2\cos(t)/\gamma_1$ and $\p_t \Up|_{y \pm 1}= -\gamma_2\sin(t)/\gamma_1$, while
\begin{eqnarray}\label{eq:rands}
r&=&H^{1/2}[(\Sr\Rey/H)^2+1]^{1/4} \cos([\tan^{-1}(\Sr\Rey/H)]/2), \\ \nonumber
s&=&H^{1/2}[(\Sr\Rey/H)^2+1]^{1/4} \sin([\tan^{-1}(\Sr\Rey/H)]/2),
\end{eqnarray}
represent the inverse boundary layer thickness and the wave number of the wall-normal oscillations of the base flow, respectively. In the hydrodynamic limit of $H \rightarrow 0$, these quantities reduce to $r=s=(\Sr\Rey/2)^{1/2}$. In the limit of $H \rightarrow \infty$, at constant $\Rey$ and $\Sr$, $r \sim  H^{1/2}$ and $s \rightarrow 0$. Note from Eq.~(\ref{eq:def_base}) that the maximum velocity over the cycle $U_0 = \max_{\{y,t\}}(\Up)=\gamma_1$ for $\Gamma \geq 1$ (henceforth, $\Gamma \geq 1$). To avoid the dependence \textcolor{\mcy}{of the maximum velocity of the steady component} on $\Gamma$, the rescaled Reynolds number $\ReyG = \gamma_1\Rey$ is defined; \textcolor{\mcy}{the use of $\ReyG$ thereby keeps the magnitude of steady inertial forces constant when varying the amplitude ratio $\Gamma$, while} in the steady limit $\ReyG \rightarrow \Rey$. The rescaled Reynolds number is used in the definitions of the critical Reynolds number ratios $\rrpc = \ReyG/\ReyCritPS$ and $\rrc = \ReyG/\ReyCritST$, relative to the critical Reynolds numbers $\ReyCritPS$ and $\ReyCritST$ for the linear instability of the pulsatile and steady base flows, respectively. \textcolor{\mcy}{Note that $\rrc$ provides a measure of $\mathit{Re}$ compared to the equivalent steady critical Reynolds number, which would otherwise vary with friction parameter $H$.} The critical Reynolds number ratio is then $\rrs=\ReyCritPS/\ReyCritST=\rrpc/\rrc$, \textcolor{\mcy}{which provides a separate measure of how effectively the pulsatile flow is able to reduce the critical Reynolds number, independent of whether turbulence can be sustained}. An example base flow, rescaled in an equivalent manner such that its maximum velocity is unity, is included in \fig~\ref{fig:prob_set}; see Ref.~\citep{Camobreco2021stability} for more.

\section{Linear transient growth} \label{sec:ltgro}

\subsection{Formulation}

The non-modal perturbation maximizing growth $G=\lVert \uvpp(t=\tau+\tzero) || / \lVert \uvpp(t=\tzero) \rVert$ in finite time interval $\tau$ from initial seed time $\tzero$ is sought. $G$ represents the gain in perturbation kinetic energy under the norm $\lVert \uvpp \rVert = \int \uvpp \bcdot \uvpp \,\mathrm{d}\Omega$ \citep{Barkley2008direct}, where $\Omega$ denotes the computational domain. $\Gmax$ is the optimal growth over the three parameters $\tzero$, $\tau$ and $\alpha$ (for a steady base flow, optimization is over $\tau$ and $\alpha$ only).

An evolution equation for the perturbations is obtained by substituting Eqs.~(\ref{eq:Q2Dc}) and (\ref{eq:Q2Dm}) into Eq.~(\ref{eq:def_pert}), and retaining only terms linear in $\uvpp$. Taking twice the curl of the result, applying the divergence free constraint, and assuming plane wave modes by virtue of the streamwise periodic domain, e.g.~$\uvpp = \uvph\exp(\ii\alpha x) + \mathrm{c.c.}$, yields the linearized perturbation evolution equation,
\begin{equation} \label{eq:ltg_forward}
\pde{\vhp}{t} = \mathcal{A}^{-1}\left[-\ii\alpha \Up \mathcal{A} + \ii\alpha \partial_{yy}(\Up) + \frac{1}{\Rey}\mathcal{A}^2  - \frac{H}{\Rey}\mathcal{A} \right]\vhp,
\end{equation}
where $\mathcal{A} = D^2-\alpha^2$ and $D$ represents $\p_y$. The adjoint evolution equation
\begin{equation} \label{eq:ltg_adjoint}
\pde{\xhp}{t} = \mathcal{A}^{-1}\left[\ii\alpha \Up \mathcal{A} + 2\ii\alpha \partial_y (\Up) D + \frac{1}{\Rey}\mathcal{A}^2  - \frac{H}{\Rey}\mathcal{A} \right]\xhp,
\end{equation}
is derived from Eq.~(\ref{eq:ltg_forward}) based on the definition of the adjoint velocity perturbation $\xhp$ introduced in Ref.~\citep{Schmid2001stability}. The domain $y \in [-1,1]$ is discretized with $\Nc+1$ Chebyshev nodes \citep{Trefethen2000spectral,Weideman2001differentiation}, while powers of the $y$-derivation operator $D$ are approximated by derivative matrices incorporating boundary conditions \citep{Trefethen2000spectral}. Boundary conditions are $\vhp = D\vhp = \xhp = D\xhp = 0$ at all walls. As the base flow is symmetric, a symmetry condition is enforced along the duct centerline, to resolve even perturbations separately, and halve the spatial resolution requirements. A third-order forward Adams--Bashforth scheme \citep{Hairer1993solving} integrates Eq.~(\ref{eq:ltg_forward}) from $t=\tzero$ to $t=\tau+\tzero$, and with `initial' condition $\xhp(\tau+\tzero)=\vhp(\tau+\tzero)$, integrates Eq.~(\ref{eq:ltg_adjoint}) from $t=\tau+\tzero$ back to $t=\tzero$. After normalizing, the next iteration proceeds, with the maximum growth then found by evaluating $\lVert \vhp(t=\tau+\tzero) \rVert$/$\lVert\vhp(\tzero)\rVert$.

The optimal growth $\Gmax$, for a given $\Rey$, $H$, $\Sr$ and $\Gamma$, is found to within fixed $5\times10^{-3}$ increments in $\alpha$, fixed $2\pi\times10^{-2}$ increments in $\tzero$, and adaptive $10^{-2}\tau$ to $10^{-3}\tau$ increments in $\tau$. For each $\alpha$ and $\tzero$, the optimization scheme employed intentionally selects too large a $\tau$ for early iterations. Once the optimal begins to converge, the target time is adaptively reduced to the time corresponding to the maximum in $\lVert\vhp\rVert_2$ over all $\tzero<t<\tau+\tzero$. After remaining in the vicinity of the local maximum, fixed adjustments to $\tau$ are performed. For each $\alpha$ and $\tzero$, the iterative scheme to find  the optimal time interval $\tauOpt$ performs anywhere from 200 to 1200 total forward-adjoint iterations, with the initial condition permitted to converge without adjusting $\tau$ for between 10 to 50 intermediate forward-adjoint iterations, depending on $\Sr$. This entire process repeats as $\alpha$ and $\tzero$ are adjusted until the optimal wave number and initial time, $\alphaOpt$ and $\tzeroOpt$, are determined. Once raw values converge, local second order polynomial fitting is applied. Spatial resolution of $\Nc=80$ for $H\leq 10$ and $\Nc=120$ for $H=100$ were selected, based on the resolution testing of Ref.~\citep{Camobreco2020transition}. Between $4\times10^5$ and $10^7$ time steps are performed for each forward or adjoint iteration \citep{Camobreco2021stability}. 

\newlength\q
\setlength\q{\dimexpr .111111111\textwidth -2\tabcolsep}
\begin{table}
\centering
\begin{tabular}{ p{0.9\q}p{0.9\q}p{0.9\q}p{0.9\q}p{0.9\q}p{0.9\q}p{1.2\q}p{1.2\q}p{1.2\q} }
\hline
$\Rey_\mathrm{B16}$ & \phantom{0}$10^3\Rey$ & \phantom{.}$10^2\Sr$ & \phantom{0}$\alphaOpt$ & \phantom{.}$\tzeroOpt$ & \phantom{0}$\tauOpt$ & \phantom{00.}$\Gmax$ & \phantom{0}$\Gmax$ \citep{Biau2016transient} & \phantom{0}$|$\% Error$|$  \\
\hline
\phantom{0}800  & \phantom{0}$7.2216$ & $7.0898$ & 7.0600 & 0.5121  & 1.7071 & $6.2841\times10^4$   & $6.2834\times10^4$   & $1.12\times10^{-2}$  \\
1000 & \phantom{0}$9.0270$ & $5.6719$ & 6.9237 & 0.4543  & 2.0584 & $3.7750\times10^6$   & $3.7743\times10^6$  & $1.98\times10^{-2}$  \\
1200 & $10.832$\phantom{0} & $4.7265$ & 6.8118 & 0.4084  & 2.3411 & $2.5699\times10^8$   & $2.5692\times10^8$   & $2.88\times10^{-2}$  \\
\hline
\end{tabular}
\caption{Validation of the linear transient growth scheme, comparing the error in the maximum growth $\Gmax$ computed herein to that of Ref.~\citep{Biau2016transient}. Note that Ref.~\citep{Biau2016transient} analyses a single parameter problem (first column). To recover purely oscillatory, hydrodynamic conditions requires $\Gamma=0$ and $H=0$. However, to match an isolated Stokes layer requires sufficiently thin boundary layers \citep{Blenn2006linear}. Thus, $\Rey$ and $\Sr$ are selected to recover the $h_\mathrm{BB06}=16$ (wall isolation) condition of Ref.~\citep{Blenn2006linear}. Overall, parameters convert as $\Rey = 2h_\mathrm{BB06}\Rey_\mathrm{B16}/(2\pi^{1/2}$), $\Sr = 2\pi^{1/2}h_\mathrm{BB06}/\Rey_\mathrm{B16}$, $\alpha = h_\mathrm{BB06}\alpha_\mathrm{B16}/\pi^{1/2}$, $\tzero=2\pi\tzeroB$ and $\tau = 2\pi\tau_\mathrm{B16}-\tzero$.}
\label{tab:tab_1}
\end{table}

To validate the forward-adjoint solver, Table~\ref{tab:tab_1} compares the fully optimized computations of Ref.~\citep{Biau2016transient} to those of the MATLAB solver detailed here. Note that Ref.~\citep{Biau2016transient} analyses a purely oscillatory hydrodynamic boundary layer flow; this was the closest setup to a pulsatile Q2D duct flow found but leaves $\Sr$ as a free parameter. Thus, to approach the results of Ref.~\citep{Biau2016transient}, the base flow must exhibit isolated boundary layers \citep{Blenn2006linear}, requiring $\Sr$ be relatively large. This induces some error, although clear agreement is still indicated in Table~\ref{tab:tab_1}.

Optimized linear transient growth computations follow. \textcolor{\mcy}{These are performed at Strouhal numbers $\Sr \lesssim 1/2$ (for the most part), to satisfy the assumption of quasi-two-dimensionality (see \S~II.1). The range of amplitude ratios $\Gamma \geq 1$ considered is based on previous results \citep{Camobreco2021stability}, as the minimum $\ReyCrit$ for $H \leq 10$ was found to be at $\Gamma \gtrsim 1.19$, while the steady limit, which appears to have the lowest $\mathit{Re}$ for sustaining turbulence of those tested, is at $\Gamma \rightarrow \infty$. However, as transient growth was exceedingly large at $\Gamma \approx 1$, results at amplitude ratios $\Gamma=100$ and $\Gamma=10$ form the majority of this work. The Reynolds number ranges focus on $\rrc \leq 1$, as reducing the Reynolds number required to sustain turbulence motivates this work. The ranges for the friction parameter $H$ were then selected to ensure that accurate double precision computations could be achieved for the corresponding ($\mathit{Sr}$, $\mathit{Re}$, $\Gamma$) combination; often this resulted in investigating lower $H$ values than are of practical interest, and even then, very long computations were required (e.g.~hundreds of forward-backward iterations with up to $10^7$ time-steps per iteration). Finally, to compare the pulsatile transient growth results to their steady counterparts} ($\Gamma \rightarrow \infty$), the following ratios are introduced: $\gst = \Gmax/\GmaxST$, $\astp=\alphaOpt/\alphaST$ and $\tst=\gamma_1\tauOpt/(\tauST\Sr)$. As $\tzeroOpt$ is arbitrary for a steady base flow, $\tzn = m+\tzeroOpt/2\pi$, for integers $m$, is defined. $\GmaxST$, $\alphaST$ and $\tauST$ were computed following Ref.~\citep{Camobreco2020transition}.

\subsection{Results: Large amplitude ratio}\label{sec:rslin}

\begin{figure}
\centering
\includegraphics[width=\textwidth]{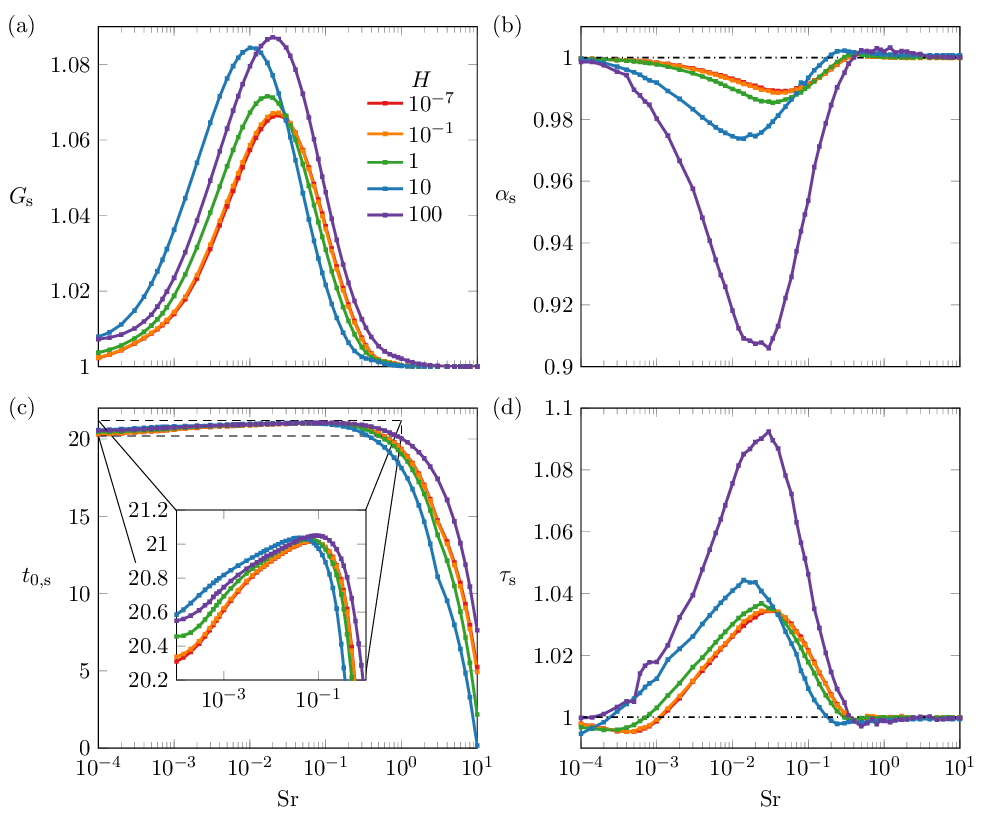}
\vspace{-12mm}
    \caption{Linear transient growth for a pulsatile base flow at $\Gamma=100$, $\rrc=1$ as a function of $\Sr$, for various $H$ (see legend). Growth was optimized over all streamwise wave numbers, initial seed times, and time intervals, with each normalized by their steady counterparts. (a) $\gst = \Gmax/\GmaxST$. (b) $\astp=\alphaOpt/\alphaST$. (c) $\tzn = m+\tzeroOpt/2\pi$ for integers $m$ (computations performed at $m=0$, with $m$ then appropriately adjusted to produce a clearer figure). (d) $\tst=\gamma_1\tauOpt/(\tauST\Sr)$.}
    \label{fig:G100_ltg}
\end{figure}

Results at $\Gamma=100$ are considered first, for a wide range of $\Sr$ at $H\leq100$. These results are all at fixed $\rrc=1$, noting that as $\rrc$ implicitly depends on $H$ through $\ReyCritST$, $\ReyG$ is adjusted accordingly. 
The maximum growth, optimized over wave numbers $\alpha$, seed times $\tzero$ and time intervals $\tau$ is depicted in \fig~\ref{fig:G100_ltg}(a), displayed as a ratio of its steady counterpart. Importantly, \fig~\ref{fig:G100_ltg}(a) reveals that $\gst\geq1$ for all $\Sr$. Thus, the optimized growth in the pulsatile system is greater than its steady counterpart at the same $\rrc$. Furthermore, the local maximum in $\gst$ consistently increases with increasing $H$, as does the sensitivity of the optimal wave number $\astp$ and optimal time interval $\tst$, shown respectively in \figs~\ref{fig:G100_ltg}(b) and \ref{fig:G100_ltg}(d). However, as $\Gamma=100$ is close to the steady limit, the percentage increases in transient growth are modest (up to 9\%). The slight shift in the $\Sr$ location of the local maximum in $\gst$ is discussed in Ref.~\citep{Camobreco2021stability}. This discussion is not recapped here, except to note that the local minima in $\rrs$ do not exactly align with the local maxima in $\gst$. Overall, the steady result is recovered quickly in the $\Sr\rightarrow\infty$ limit (oscillating boundary layers of immaterial thickness), while the steady limit is only slowly being recovered as $\Sr\rightarrow0$ (viscous diffusion having annihilated all inflection points over the entire cycle).

\begin{figure}
\centering
\includegraphics[width=\textwidth]{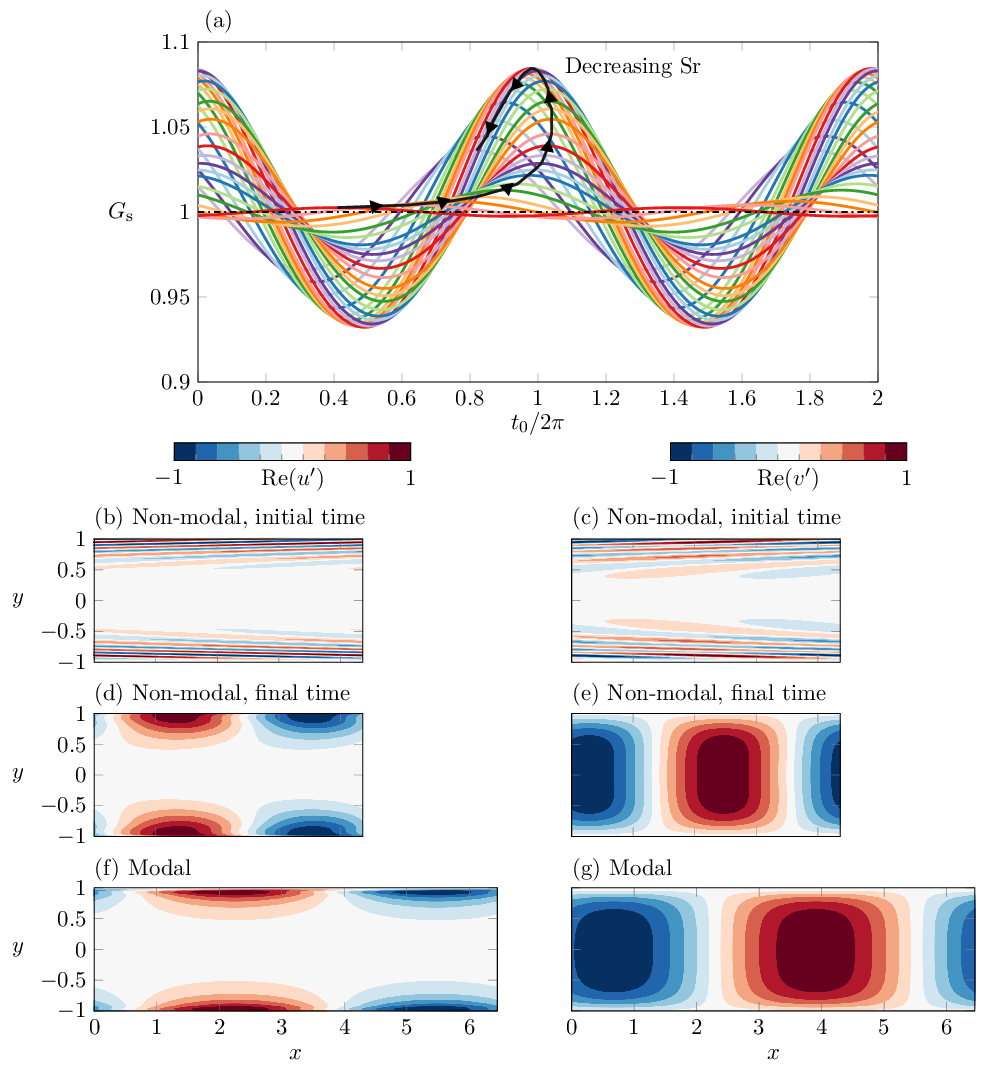}
\vspace{-12mm}
    \caption{(a) Linear transient growth for a pulsatile base flow at $\Gamma=100$, $\rrc=1$, $H=10$ at various $10^{-3} \leq \Sr \leq 0.3$, normalized by the equivalent growth for a steady base flow. At fixed $\alpha = \alphaOpt$ for each $\Sr$, growth was optimized over all time intervals for each initial time $0<\tzero/2\pi<1$ (the data set is replicated over an additional period for clarity). The maximum growth over all $\Sr$ occurred at $\Sr=10^{-2}$ \textcolor{\mcy}{with $\tzero/2\pi=0.98$, for which the optimal transient growth modes at $\tzeroOpt$ are shown in (b, c), evolved to $\tzeroOpt+\tauOpt$ in (d, e), and for which the corresponding modal instability is shown in (f, g), all normalized to unit maximum. Real parts of streamwise velocity: (b,d, f). Real parts of wall-normal velocity: (c, e, g).}}
    \label{fig:G100_sweep}
\end{figure}

Tuning $\tzero$ for each $\Sr$ was shown to be key to generating large modal intracyclic growth \citep{Camobreco2021stability}, by ensuring perturbation energy growth coincided with the deceleration of the base flow. It is of interest whether tuning $\tzero$ for non-modal perturbations is equally important when obtaining $\Gmax$, having had to optimize for not only $\tzeroOpt$, but also $\alphaOpt$ and $\tauOpt$. To determine the importance of $\tzero$ for non-modal growth, the relevance of the optimal time interval $\tauOpt$ must be untangled. At $\Gamma=100$, the intracyclic growth is order unity over a wide range of $\Sr$ \citep{Camobreco2021stability}. Thus, very little is lost in evolving over the acceleration phase of the base flow, and so the time interval for transient growth need not be the order of the deceleration phase ($\tau \approx \pi$). As transient growth of the steady base flow is order 100 (at $H=10$), there is much to benefit by instead selecting a time interval $\tau/\tauST \approx 1$ at $\Gamma = 100$. Even in spite of this reduced relevance of $\tau$ at $\Gamma = 100$ (unlike at $\Gamma = 10$, cf.~\figs~\ref{fig:G100_ltg}d and \ref{fig:G10_ltg}d), the tuning of $\tzn$ is still important (at both $\Gamma = 100$ and $\Gamma = 10$, cf.~\figs~\ref{fig:G100_ltg}c and \ref{fig:G10_ltg}c). To further analyse the role of $\tzero$, \fig~\ref{fig:G100_sweep}(a) depicts growth for $0 \leq \tzero \leq 2\pi$ having optimized $\tau$ for each $\tzero$, with the data set replicated to span $0 \leq \tzero \leq 4\pi$ for clarity. Even though $\tau$ is optimized for each $\tzero$ separately, some $\tzero$ have $G/\GmaxST<1$ for each $\Sr$. Thus, although the addition of a pulsatile component was shown to increase transient growth (relative to the steady counterpart) in \fig~\ref{fig:G100_ltg}(a), this is only true for some $\tzero$, such that pulsatility can be detrimental for poorly selected $\tzero$, even when optimizing $\tau$. Varying $\Sr$ then varies the potential for adjustments of $\tzero$ to lead to improved $\gst$.

\textcolor{\mcy}{The resemblance of the optimal modes to their steady counterparts is also briefly considered for the initial- and final-time optimized perturbation in \figs~\ref{fig:G100_ltg}(b) through (e). As shown, the initial condition optimizing growth for the pulsatile base flow appears as a series of backwards-leaning perturbations, \figs~\ref{fig:G100_ltg}(b, c), typical of the Orr mechanism for steady flows \citep{Orr1907stability,Jimenez2013linear}. As the initial perturbations tilt into the mean shear, they grow, and rapidly form a duct-spanning Tollmien--Schlichting wave at the optimal time, \figs~\ref{fig:G100_ltg}(d, e). The differences between the transiently-generated TS wave and the modal (intracyclic) equivalent from the linear stability analysis are further shown to be minimal by comparison with \figs~\ref{fig:G100_ltg}(f, g), in which the leading eigenmode is plotted. Note that there is a slight difference in the wavenumber of the nonmodal ($\alpha \approx 1.44$) and modal ($\alpha \approx 0.96$) perturbations, to maximize transient growth in the former case, and to minimize exponential decay in the latter, but both otherwise appear as duct-spanning TS waves.}

\subsection{Results: Intermediate amplitude ratio}\label{sec:rslini}

\begin{figure}
\centering
\includegraphics[]{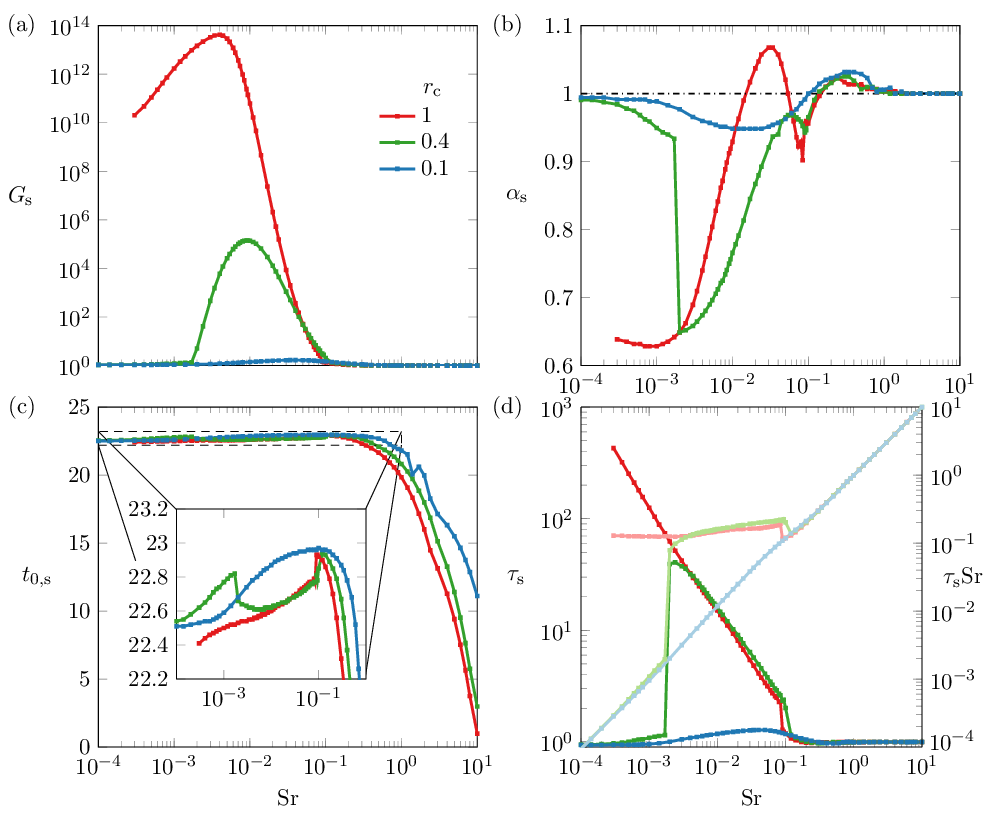}
\vspace{-12mm}
\caption{Linear transient growth for a pulsatile base flow at $\Gamma=10$ and $H=10$, as a function of $\Sr$, for various $\rrc$ (see legend). Growth was optimized over all streamwise wave numbers, initial seed times, and time intervals, with each normalized by their steady counterparts. (a) $\gst = \Gmax/\GmaxST$. (b) $\astp=\alphaOpt/\alphaST$. (c) $\tzn = m+\tzeroOpt/2\pi$ for integers $m$ (a clearer figure is produced when $m$ is appropriately adjusted). (d) $\tst=\gamma_1\tauOpt/(\tauST\Sr)$ (left-axis, darker lines), $\tst \Sr=\gamma_1\tauOpt/\tauST$ (right-axis, ligher lines).}
    \label{fig:G10_ltg}
\end{figure}

\Fig~\ref{fig:G10_ltg} depicts optimized \textcolor{\mcy}{transient growth} parameters at the amplitude ratio $\Gamma=10$. Even at this intermediate $\Gamma$, the optimised growth relative to \textcolor{\mcy}{its} steady counterpart \textcolor{\mcy}{is} already exceedingly large at intermediate $\Sr$, peaking at $\gst \approx 10^{14}$ at $\rrc=1$, \fig~\ref{fig:G10_ltg}(a). Whether such large transient growth can prove advantageous for inciting transitions to turbulence, when occurring at a wavenumber similar to the steady counterpart, \fig~\ref{fig:G10_ltg}(b), will be considered further in the following section. While there are otherwise some similarities in the general trends for $\gst$, $\astp$ and $\tzn$ at $\Gamma=10$ and $\Gamma=100$, there is one striking difference. \textcolor{\mcy}{T}aking $\rrc=0.4$ as an example, \textcolor{\mcy}{there is a large increase in $\gst$ at intermediate $\Sr$,} with discontinuities present in the variations of $\astp$ and $\tzn$ at $\Sr$ of $\approx 10^{-1}$ and $\approx 10^{-3}$ at $\Gamma=10$. \textcolor{\mcy}{However, across these discontinuities, the initial conditions optimising growth still appear similar to the Tollmien--Schlichting waves presented in \fig~\ref{fig:G100_sweep}. Instead, the presence of the discontinuities is related to: the variations in the base flow as a function of $\Sr$, the comparison of optimized growth in \fig~\ref{fig:G10_ltg} at constant $\rrc$ (constant $\Rey$), rather than at constant $\rrpc$ ($\Rey$ varying with $\Sr$ to ensure constant criticality of the pulsatile flow), and finally, as a fixed $\rrc$ comparison mixes exponential and transient growth whenever $\rrpc  > 1$ (supercritical conditions), although this latter issue is accounted for in \fig~\ref{fig:TGonIC}. These factors aside,} at \textcolor{\mcy}{fixed $\rrc$} and either small or large $\Sr$, \textcolor{\mcy}{the initial condition, and transient growth, are dominated by steady mechanisms; both small and large $\Sr$ recover the steady result}, either in the diffusive limit, or in the non-penetration limit of infinitesimally thin oscillating boundary layers, respectively. However, for intermediate $\Sr$, \textcolor{\mcy}{transient inertial forces begin to outweigh viscous ones, enabling} the optimal perturbation \textcolor{\mcy}{to generate} orders of magnitude more growth. These \textcolor{\mcy}{differences are further considered in} \fig~\ref{fig:G10_ltg}(d), which shows the optimal time interval, relative to the steady counterpart, as a function of $\Sr$. At large $\Sr$, the \textcolor{\mcy}{pulsatile} optimal matches the optimal of the steady flow\textcolor{\mcy}{, and so $\tst \approx 1$}. With reducing $\Sr$, $\tst$ remains approximately unity until $\Sr \approx 10^{-1}$. There, after a discontinuity (for larger $\rrc$), $\tst$ increases almost linearly with $\Sr$, with increasingly large time intervals relative to the steady equivalent. Within the range $10^{-3}\lesssim\Sr\lesssim 10^{-1}$, these optimal time intervals, replotted as $\tst\Sr$, appear roughly constant. This more truly pulsatile mode, present for $10^{-3}\lesssim\Sr\lesssim 10^{-1}$, takes advantage of the effective increase in the duration of the acceleration and deceleration phases of the base flow, while inflection points are present, to \textcolor{\mcy}{achieve significantly more transient growth. Once} $\Sr$ becomes too small, $\Sr \lesssim 10^{-3}$, the $\tst$ curve again shows a discontinuity, falling back to the optimal for the steady base flow, which only exhibits modest $\Gmax$. Note that this discontinuity \textcolor{\mcy}{(in $\tst$)} could not be observed at $\rrc=1$, as the convergence of the computations became quite problematic for $\Sr \lesssim 2\times10^{-4}$. \textcolor{\mcy}{Overall, a range of frequencies $10^{-3}\lesssim\Sr\lesssim 10^{-1}$ have been identified, at fixed sufficiently large Reynolds numbers and amplitude ratios, at which growth can be greatly enhanced by transient inertial mechanisms. The question is then whether this growth is of a modal or non-modal origin.}

\subsection{Non-modal vs.~modal intracyclic growth}

\begin{figure}
\centering
\includegraphics[]{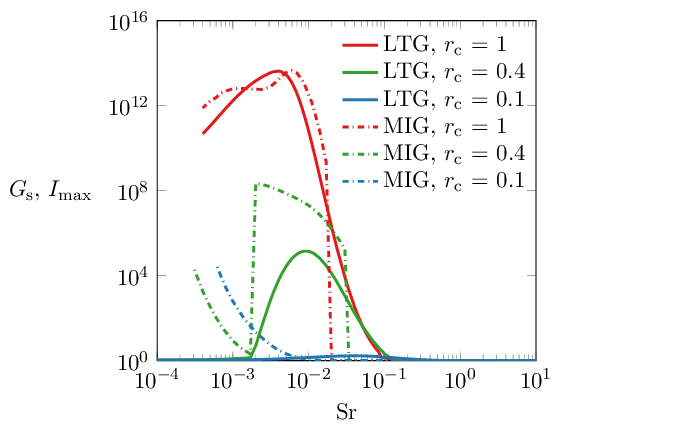}
\vspace{-5mm}
    \caption{Comparison between the optimized linear transient growth $\gst = \Gmax/\GmaxST$ normalized by the steady component (denoted LTG) and the intracyclic growth $\IC$ at identical parameters, measured as the ratio of the maximum to the minimum energy over the cycle, after factoring out exponential growth or decay  (denoted MIG). Ultimately, the transient growth is predominantly comprised by the modal growth (the steady contributions are order $10^2$ at $\rrc=1$). Note that for large $\Sr$, once $\IC$ was unity to within $\pm 10^{-8}$, the computations were ceased and a unit value plotted to continue the trends. Similarly for $\Sr \lesssim 8\times10^{-4}$, the computations became untenable due to the relative level of exponential growth or decay.}
    \label{fig:TGonIC}
\end{figure}

It has yet to be demonstrated that pulsatile mechanisms can provide significant non-modal growth potential, and take advantage of non-normality. Rather, in the intermediate frequency regime, the total growth has previously been shown to predominantly consist of modal intracyclic growth \citep{Pier2021optimal}. It is \textcolor{\mcy}{therefore} of interest to compare this broader investigation on the role\textcolor{\mcy}{s} of $\Sr$ \textcolor{\mcy}{and $\Rey$ at an intermediate amplitude ratio $\Gamma=10$} to those of Refs.~\citep{Pier2021optimal,Kern2021transient,Kern2022subharmonic}. As presented in \fig~\ref{fig:G10_ltg}, similar conclusions are drawn in this work, despite the considerably higher Reynolds numbers (order $10^4$) associated with these quasi-two-dimensional base flows. The total transient growth is presented as $\gst = \Gmax/\GmaxST$, identical to figure \fig~\ref{fig:G10_ltg}(a), so as to remove the steady contribution from the comparison. Whether the remaining growth is then generated by modal mechanisms is assessed by computing the intracyclic growth $\IC$ at identical conditions to that at which the transient growth computations were performed (identical $\Gamma$, $\Sr$, $\Rey$, $H$, $\alpha$).  The intracyclic growth is computed by first determining the leading eigenmode, by direct forward evolution of Eq.~(\ref{eq:ltg_forward}), with renormalization to unit norm at $2\pi$ intervals, and then taking the ratio of the maximum to the minimum perturbation kinetic energy attained over the cycle. The perturbation energy growth is computed in the standard energy norm \citep{Schmid2001stability} as $\lVert \uvph(t) \rVert_E$, where $\uhp = \ii D \vhp/\alpha$, with Clenshaw--Curtis integration over the computational domain \citep{Trefethen1993hydrodynamic}. As $\rrc=1$ is not equivalent to $\rrpc=1$, some exponential growth or decay may still be occurring, so $\IC$ is computed as in \citep{Pier2021optimal}, taking the ratio of the maximum to minimum perturbation energy over the cycle, after factoring out any exponential growth or decay. For the transient growth, thanks to the adjoint formulation, computing $\lVert \vhp(t=\tau+\tzero) \rVert_2/\lVert\vhp(\tzero)\rVert_2$ is equivalent to $\lVert \uvph(t=\tau+\tzero) \rVert_E/\lVert \uvph(t=\tzero) \rVert_E$ at the target time.

From \fig~\ref{fig:TGonIC}, it is evident that at $\rrc = 1$, the intracyclic growth provides almost the entirety of the remaining non-modal transient growth (the magnitudes of the solid and dot-dashed curves are similar), and furthermore dwarfs the steady contribution (order $10^{12}$ compared to order $10^2$). Thus, no new pulsatile mechanisms able to take advantage of non-normality have been observed. At lower $\rrc$, it even appears that the intracyclic growth is larger than the non-modal growth. However, recall that $\IC$ is computed after factoring out exponential growth or decay, which would still be measured in $\gst$. Thus, when $\rrc$ (and thereby $\rrpc$) is far from unity, and particularly as $\Sr \rightarrow 0$ when exponential decay is exceedingly rapid, $\IC$ can appear much larger than $\gst$, although such $\IC$ are not realisable in the face of such decay. Ultimately, whenever the total non-modal growth is large, the predominant contribution is from the modal intracyclic growth, even at the larger Reynolds numbers considered herein. Thus, optimizing the initial condition is unlikely to provide a significant benefit, with white noise initial conditions thereby equally efficient at generating amplified perturbations.



\section{Nonlinear growth and transitions to turbulence} \label{sec:nlinm}

Much of the preceding section focused on establishing the parameter ranges within which non-modal growth was large, and determining the overall optimal parameters. However, the overarching aim of the nonlinear investigations in this work is to identify whether turbulence can be triggered and sustained at lower $\rrc$ than for a purely steady base flow (i.e.\ at $\rrc\lesssim0.78$), regardless of the initial condition. That $\rrc\gtrsim0.78$ is required to sustain turbulence for a steady Q2D base flow at $H=10$ is based on simulations at $\rrc = 0.7$ and $\rrc = 0.8$ from Appendix G of Ref.~\citep{Camobreco2021thesis}, where only the latter $\rrc = 0.8$ sustains turbulence, as well as additional unpublished simulations in smaller $\rrc$ increments. Notably, this $\rrc$ requirement will be more than sufficient for white noise initial conditions to be almost equally as useful as their optimized counterparts.

\subsection{Numerical methods}

Direct numerical simulation of Eqs.~(\ref{eq:Q2Dc}) and (\ref{eq:Q2Dm}) are performed with an in-house spectral element solver, employing a third order backward differencing scheme, with operator splitting, for time integration. High-order Neumann pressure boundary conditions are imposed on the oscillating walls to maintain third order time accuracy \citep{Karniadakis1991high}. The Cartesian domain is discretized with quadrilateral elements over which Gauss--Legendre--Lobatto nodes are placed. The mesh design, and validation of the solver including both friction and pulsatility, have been previously discussed  \citep{Camobreco2020transition,Camobreco2021stability,Camobreco2021weakly}. There are 48 spectral elements in the wall-normal direction, while simulations at $\Gamma=10$ and $\Gamma = 1.19$ respectively employed $12$ and $48$ spectral elements in the streamwise direction. Each spectral element has polynomial basis $\Np=19$. These resolution requirements are generally dictated by the Reynolds number and the overall energetics of the turbulence; the transverse length scales of both the steady Shercliff boundary layers and the oscillating boundary layers were large (of the order of the duct half-height), for most of the parameter combinations ($H$, $\Sr$) investigated, and particularly those maximizing intracyclic growth (intermediate $\Sr$).

The initial condition $\uvp(t=\tzero)=\Uvp+\uvpp$ is the sum of the laminar profile, Eq.~(\ref{eq:def_base}), and a perturbation either composed of a linear transient growth optimal (recomputed in the in-house solver), or white noise. Perturbations have specified initial energy $E_{0}(t=\tzero)=\int\upp^2+\vpp^2\,\mathrm{d}\Omega/\int \Up^2(t=\tzero)\,\mathrm{d}\Omega$, where $\Omega$ represents the computational domain. For white noise $\tzero$ is not relevant, as seeding at an inefficient $\tzero$ does not harm the (lack of) structure of the initial condition. The flow is driven by a homogeneous pressure gradient, $\partial \Pp/\partial x = \gamma_1 H\cosh(H^{1/2})/[\cosh(H^{1/2})-1]\Rey$, with the pressure decomposed into a linearly varying and fluctuating periodic component, as $\pp = \Pp + \ppp$, respectively. Periodic boundary conditions $\uvp(x=0)=\uvp(x=W)$ and $\ppp(x=0)=\ppp(x=W)$ are applied at the downstream and upstream boundaries. The domain length $W=2\pi/\alpha$ is set to match either the wave number minimizing the decay rate of the leading eigenmode $\amax$ or that optimizing linear transient growth $\alphaOpt$. Synchronous lateral wall movement generates the oscillating flow component, with boundary conditions $\Up(y \pm 1,t) = \gamma_2 \cos(t)$.


\subsection{Results}

\begin{figure}
\centering
\includegraphics[width=\textwidth]{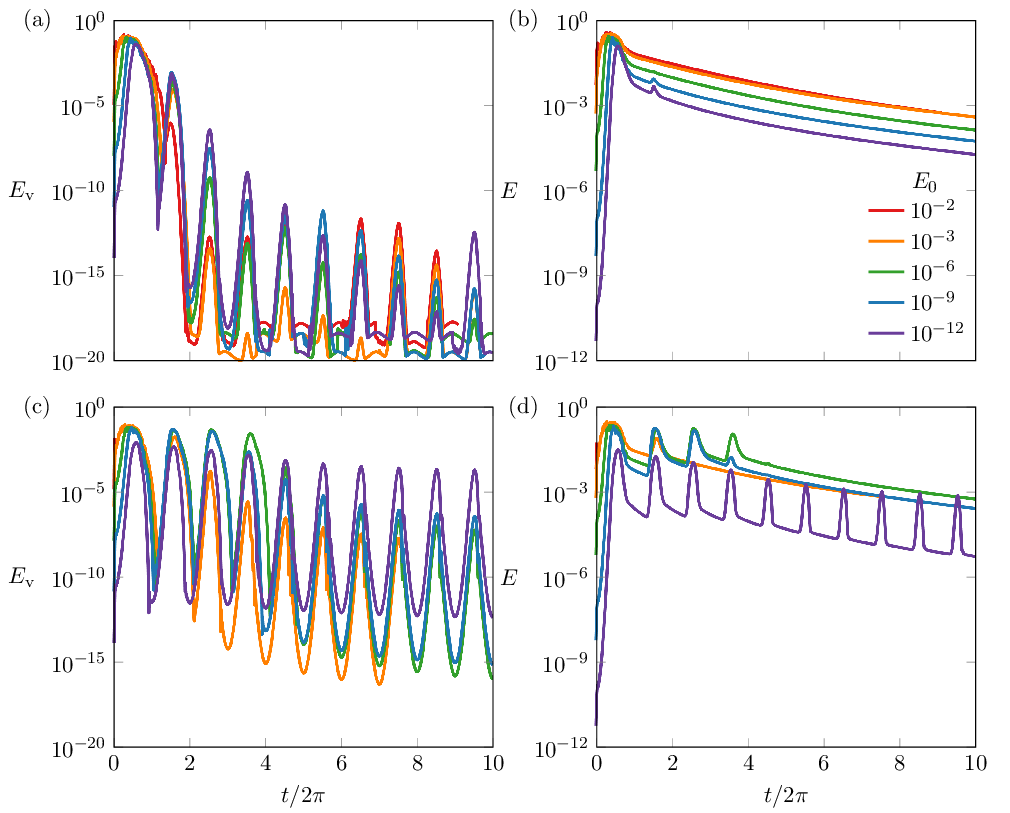}
\vspace{-12mm}
    \caption{Nonlinear evolution of linearly optimized initial conditions at $H=10$, $\Gamma=10$, $\Sr=9\times10^{-3}$ and $\rrc=0.737$ ($\rrpc=1$) for various initial perturbation energies $\Ezero$. This $\Sr$ yields the minimum $\rrs$ at this $H$ \citep{Camobreco2021stability}. Both cases optimized over all initial conditions, initial times and final times. (a--b) Streamwise wavenumber $\alphaOpt$, maximizing transient growth, but weakly subcritical due to the shift in wavenumber. (c--d) Streamwise wavenumber $\amax$, neutrally stable. (a \& c) $\Ev = \int\vpp^2\,\mathrm{d}\Omega$. (b \& d) $E = \int\upp^2+\vpp^2\,\mathrm{d}\Omega$.}
    \label{fig:G10DNS}
\end{figure}

The first of the nonlinear simulations presented, in \fig\ \ref{fig:G10DNS}, are at $\Gamma=10$, $H=10$ and specifically $\Sr=9\times10^{-3}$, so that $\rrpc=1$ and $\rrc=0.737<0.78$, \emph{i.e.}~less than the minimum $\rrc$ for a steady base flow to sustain turbulence. Thus, if turbulence is sustained (this being the aim), an oscillatory driving force would provide a clear advantage over a steady one. However, regardless of the streamwise wavenumber or initial perturbation energy, turbulence is hardly triggered, let alone sustained, even with large $\Ezero$. This result is similar to that of a previous Q2D investigation with white noise initial conditions when intracyclic growth is large \citep{Camobreco2021stability}. Perturbations over a wide range of initial energies rapidly grow, until the base flow becomes nonlinearly modulated. After a few periods, the flow settles, with either fluctuations about a constant energy or a slowly decaying modulation. These large amplitude fluctuations are clearer when considering the wall-normal perturbation energy component (\figs\ \ref{fig:G10DNS}a and c).


Thus, even at a Reynolds number only slightly below $\rrc=0.78$, and in spite of a massive increase in the transient growth potential (although fed by a modal mechanism), turbulence is neither clearly triggered, nor sustained. Previous low $\Rey$ investigations at low $\Gamma$ (order 1) were also unsuccessful at sustaining turbulence \citep{Camobreco2021stability}, but \textcolor{\mcy}{even the much higher $\Rey$ tested herein} at intermediate amplitude ratios (order 10, much closer to the steady limit) provided little improvement when it came to sustaining turbulence. Ultimately, the degree of subcriticality (22\%) at which turbulence can be sustained in a steady base flow is not carried across to pulsatile flows at $\Gamma=10$ (for any $\Sr$ tested), with pulsatile flows unable to sustain turbulence even at critical conditions ($\rrpc=1$), \textcolor{\mcy}{such that the steady base flow sustains turbulence at lower Reynolds numbers than its pulsatile counterpart.}

\textcolor{\mcy}{For the pulsatile base flow, it} is thus unclear what $\Gamma$ is best; it may well be the case that $\Gamma\rightarrow\infty$ (a steady base flow) is optimal for sustaining turbulence. Note that the critical Reynolds number rapidly recovers the value for the steady flow with increasing $\Gamma$ ($\ReyCritPS \rightarrow \ReyCritST$), e.g.~at $\Gamma=100$, the critical Reynolds number $\ReyCritPS$ is within $\approx 1$\% of $\ReyCritST$ \citep{Camobreco2021stability}. Any benefits in reducing $\ReyCrit$, if this is even a viable strategy to help sustain turbulence at lower $\Rey$, thereby become minimal. Thus, simulations of the converse, a smaller amplitude ratio ($\Gamma=1.19$) are revisited herein in supercritical conditions $\rrpc>1$, while still keeping $\rrc<0.78$. At $H=10$, $\Gamma=1.19$ provides the minimum $\rrs$ over all $\Gamma$, at $\Sr=5.6\times10^{-3}$ \citep{Camobreco2021stability}. At this small an amplitude ratio, optimized transient growth is of order $10^{60}$ at $\rrc=0.4$ and well over order $10^{100}$ at $\rrc=1$. Due to this, white noise initial conditions are employed, as the transient growth scheme struggles to accurately converge. Furthermore, such conditions would generate a total non-modal growth composed almost entirely of the intracyclic (modal) contribution, as discussed in the preceding section.

\begin{figure}
\centering
\includegraphics[width=\textwidth]{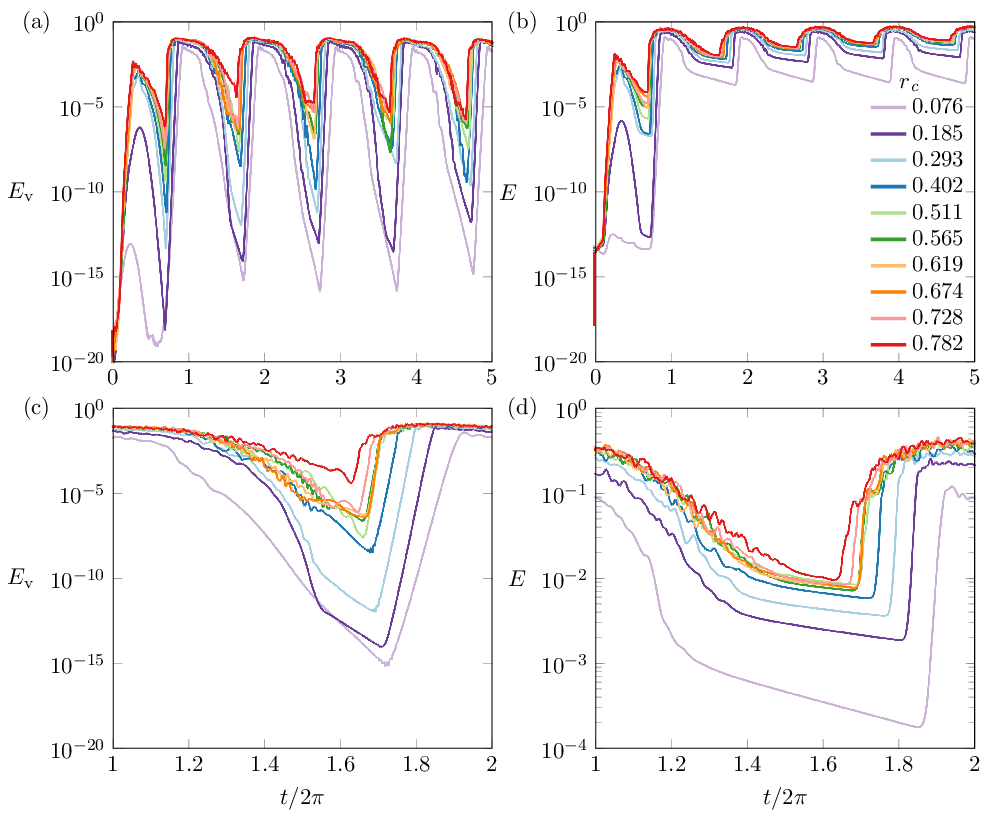}
\vspace{-12mm}
    \caption{Nonlinear evolution of white noise initial conditions ($\Ezero=10^{-18}$) at $H=10$, $\Gamma=1.19$ and $\Sr=5.4\times10^{-3}$ for a wide range of $\rrc \lesssim 0.78$, i.e.~approximately up to the lowest $\rrc$ for which a steady base flow sustains turbulence at this $H$. (a) $\Ev = \int\vpp^2\,\mathrm{d}\Omega$. (b) $E = \int\upp^2+\vpp^2\,\mathrm{d}\Omega$. (c) Zoom-in of (a) over one period. (d) Zoom-in of (b) over one period.}
    \label{fig:G119DNS}
\end{figure}

\Fig\ \ref{fig:G119DNS} depicts both subcritical ($\rrpc \lesssim 1$) and supercritical ($\rrpc \gtrsim 1$) white noise simulations in the range $0.078\lesssim\rrc\lesssim 0.78$ at $\Gamma=1.19$. Even over this large a range of $\rrc$, including supercritical $\rrpc \gtrsim 1$, turbulence does not appear to be sustained, with fluctuations dying out over $1.5 \lesssim t/2\pi \lesssim 1.6$ (\figs\ \ref{fig:G119DNS}c and d), and similarly for each cycle thereafter, particularly at lower $\rrc$. At larger $\rrc$, the fluctuation energy during these partial relaminarization events still reduces by over three orders of magnitude relative to the turbulent portions of the cycle, although weak fluctuations may remain. Thus, the ultimate bound of $\rrc>0.78$ as the minimum Reynolds number to sustain turbulence still stands, and is provided by a steady base flow. No pulsatile base flows (of those tested), were able to sustain turbulence over the entire cycle at an equivalent or lower $\Rey$. Furthermore, except for the lowest few $\rrc$ tested, there appears to be little difference in the maximum energy attained each pulsation cycle at $\Gamma=1.19$. At best, larger $\rrc$ appear to maintain a higher energy state for a greater fraction of the pulsation period, with less decay and earlier growth within each cycle. As observed in previous investigations at lower $\rrc$ \citep{Camobreco2021stability}, a nonlinear base flow modulation forms, indicated by the arrested decay of the perturbation energy $E$ after the first pulsation cycle. While the propensity for exceedingly large growth is present at small $\Gamma$, the severe decay each cycle hampers the ability to efficiently sustain turbulence at a Reynolds number lower than that for a steady base flow.

\subsection{Characteristics of the turbulent episodes}

\begin{figure}
\centering
\includegraphics[width=\textwidth]{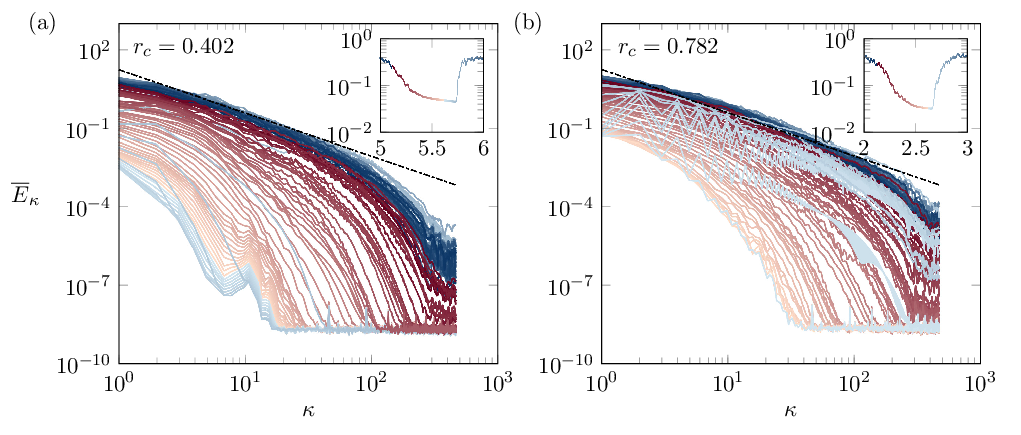}
\vspace{-12mm}
    \caption{Fourier spectra of the perturbation energy ($y$-averaged) at equispaced snapshots over one cycle at (a) $\rrc = 0.402$ and (b) $\rrc = 0.782$; $H=10$, $\Gamma=1.19$ and $\Sr=5.4\times10^{-3}$. The black dot-dashed line denotes a $\kappa^{-5/3}$ trend.  For $\rrc=0.402$ the spectra are plotted from $5 \lesssim t/2\pi \lesssim 6$, and for $\rrc = 0.782$ from $2 \lesssim t/2\pi  \lesssim 3$. Inset figures are the corresponding perturbation energies $E = \int\upp^2+\vpp^2\,\mathrm{d}\Omega$ over one cycle. Dark red to light red is through the acceleration phase, light blue to dark blue through the deceleration phase.}
    \label{fig:FoDNS}
\end{figure}

To further highlight the temporally intermittent nature of the turbulence, \fig~\ref{fig:FoDNS} depicts $y$-averaged Fourier spectra $\overline{E}_\kappa$ at equispaced temporal snapshots, where $\hat{E}_\kappa(y) = N_f^{-1}\sum_{n=0}^{N_f-1}|\uvpp|^2\exp(-2\pi \mathrm{i}\kappa n/N_f)$, and where the $n^\mathrm{th}$ coefficient $x_n=2\pi n/(\alpha N_f)$ spans the streamwise-periodic domain. The number of interpolation points in $x$ is $N_f=10000$, and 21 equispaced $y$ slices are averaged across. From these spectra, a turbulent state is apparent even at $\rrc = 0.402$ for certain phases of the cycle, as shown in \fig~\ref{fig:FoDNS}(a), at an $\rrc$ well below that required for a steady base flow to indefinitely sustain turbulence. Specifically, turbulence is observed predominantly during the deceleration phase of the base flow ($0 \lesssim t_s/2\pi\lesssim 0.25$ and $0.75 \lesssim t_s/2\pi\lesssim 1$), when an adverse pressure gradient is effectively present. However, throughout the start of the acceleration phase (from $t_s/2\pi\approx 0.25$), the turbulence begins to collapse, and a lower-energy (relaminarizing) state is attained throughout the latter parts of the acceleration phase. Toward the end of the acceleration phase (around $t_s/2\pi\approx 0.75$), a transition to turbulence is again observed in the spectra. At a much larger $\rrc = 0.782$, as shown in \fig~\ref{fig:FoDNS}(b), the relaminarization occurs later in each period, and the relaminarization does not progress as far in the acceleration phase before transition re-occurs. Note that a floor in the energy spectra is present for $\kappa \gtrsim 30$ (at $\overline{E}_\kappa \approx 10^{-9}$ for both $\rrc$), which is a clear indication of this partial relaminarization. This floor in the spectra is not present for a steady base flow that produces temporally intermittent turbulence at a similar $\rrc$ \citep{Camobreco2020transition}. Thus, pulsatility still acts to delay transition to sustained turbulence even at larger $\rrc$, with the partial relaminarization more effective, albeit still incomplete, than the partial relaminarization observed in intermittent (non-fully-developed) turbulence produced by a steady base flow. These relaminarization events are reflected in \fig\ \ref{fig:G119DNS}(c), as the wall-normal perturbation kinetic energy $E_v$ drops by about 8 orders of magnitude by $(t-2\pi)/2\pi \approx 0.65$ at $\rrc = 0.402$ and by about 3 orders of magnitude by $(t-2\pi)/2\pi \approx 0.6$ at $\rrc = 0.782$. \textcolor{\mcy}{Note that at $\rrc = 0.782$, this 3 order of magnitude reduction in $E_v$ is not sufficient to recover linear dynamics throughout the deceleration phase, as both $E$ and $E_v$ still remain relatively large.}

\begin{figure}
\centering
\includegraphics[width=0.96\textwidth]{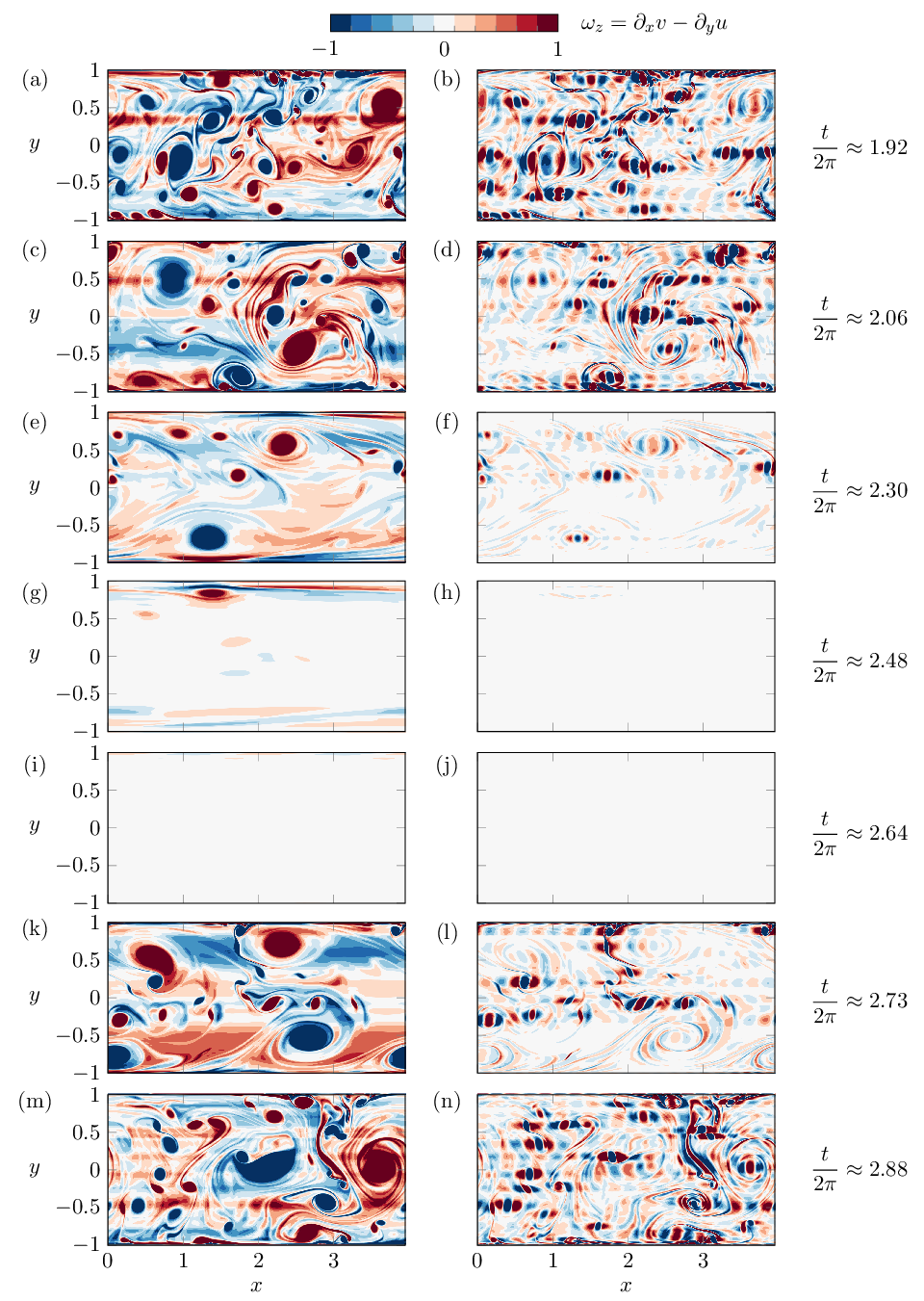}
\vspace{-7mm}
    \caption{Vorticity snapshots over one cycle at $\rrc = 0.782$, $H=10$, $\Gamma=1.19$ and $\Sr=5.4\times10^{-3}$. Left column: zeroth-mode streamwise high-pass filtered (retaining $\hat{\omega}_{z,|\kappa| \geq 1}$). Right column: zeroth-through-ninth mode streamwise high-pass filtered (retaining $\hat{\omega}_{z,|\kappa| \geq 10}$).}
    \label{fig:snapshots}
\end{figure}

Finally, to visualize the dynamics throughout the cycle at $\rrc = 0.782$, \fig\ \ref{fig:snapshots} presents snapshots of the transverse vorticity $\omega_z = \partial_x v - \partial_y u$, over the same period as \fig\ \ref{fig:FoDNS}(b). A wide range of scales are visible in the contours of the vorticity perturbation (left column of \fig\ \ref{fig:snapshots}), with the smaller scales highlighted with a high-pass filter, retaining streamwise modes with $|\kappa| \geq 10$ (right column of \fig\ \ref{fig:snapshots}). In turbulent portions of the cycle, predominantly during the deceleration phase, both thin shear layers and intense vortical regions are visible ($2\lesssim t/2\pi \lesssim 2.25$). Throughout the start of the acceleration phase, the smaller scale modes decay, and only thin, highly streamwise-sheared structures remain during the partial-relaminarization ($t/2\pi \approx 2.5$), which gravitate toward the walls. Throughout the end of the acceleration phase, and start of the deceleration phase ($t/2\pi \approx 2.75$), turbulence is reinvigorated, occupying the entire wall-normal extent of the duct (both the Shercliff and oscillating boundary layers are of the order of the duct half-height in thickness at $H=10$ and $\Sr=5.4\times10^{-3}$), with a preponderance of small scale structures reappearing. \textcolor{\mcy}{That these concentrated, highly vortical structures appear haphazardly throughout the duct} is in contrast to subcritical turbulence for a steady base flow, where the turbulent structures typically appear more wall-bounded at a similar $\rrc$ \citep{Camobreco2021weakly} or a much larger $\rrc$ \citep{Falkovich2018turbulence}. This may also be related to the route to turbulence. For the steady base flow, transition occurred via an edge state related to a nonlinear Tollmien--Schlichting wave \textcolor{\mcy}{which spanned the duct-height} \citep{Camobreco2021weakly}, and which the turbulent dynamics still revolved around, whereas turbulence is reinvigorated each cycle from small perturbation amplitudes for these pulsatile flows \textcolor{\mcy}{(\fig\ \ref{fig:snapshots}i to \ref{fig:snapshots}k)}.

\section{Conclusions} \label{sec:concl}

This work had two aims. First, assess whether non-modal transient growth mechanisms in quasi-two-dimensional flows were \textcolor{\mcy}{enhanced} by pulsatility. Second, determine whether turbulence in pulsatile Q2D flows could be sustained at lower Reynolds numbers than for steady flows. As to the former aim, non-modal mechanisms were not enhanced by pulsatility, with the total growth at intermediate frequencies almost entirely consisting of the combination of modal growth from the oscillatory component, and non-modal growth from the steady component. This was in spite of the larger Reynolds numbers investigated in this work. Thus, initial conditions optimized to attain large transient growth were of little benefit at intermediate frequencies, as the modal intracyclic growth dwarfed the steady transient growth contribution, and often proved sufficient to propel low amplitude white noise initial conditions to energies at which nonlinearity becomes important (unlike for steady base flows, which would require initial conditions with much higher energies). Transient growth of the steady base flow component was only relevant at either much higher or lower frequencies, which were therefore not of great interest.

As to the latter aim, even though large transient growth could be obtained by modal mechanisms, turbulence was never sustained at a Reynolds number lower than that for a steady base flow. Although turbulence could often be triggered at very low Reynolds numbers, turbulence was not sustained past the severe decay induced during the acceleration phases of the pulsatile base flow, as highlighted in the Fourier spectra of the perturbation kinetic energy. It thus appears that the most efficient means of sustaining turbulence in Q2D flows is with a steady driving force, since pulsatility delays the transition to sustained turbulence in Q2D flows. On the other hand, while a prime objective of this work was to find ways to promote turbulence with pulsatility, it turns out that pulsatility instead offers an effective way to partially relaminarize the flow. This is consistent with the ability for pulsatility to disrupt sustained turbulence by inducing quasi-laminar flow states which reduce drag, and thereby pumping costs, as recently shown for non-MHD flows \citep{Rota2023on}. 

\textcolor{\mcy}{As to the robustness of these Q2D results, it is admitted that quasi-two-dimensionality would only be well satisfied at scales larger than the thickness of the Shercliff layers (of thickness $L/10^{1/2} \sim 0.3L$, i.e.~30\% the duct half-height). Turbulent scales smaller than this would become increasingly three-dimensional. By the same token, 3-D instability mechanisms could independently trigger a subcritical 3-D transition, so long as the instabilities remain physically small, while any larger scales they generate would rapidly tend to quasi-two-dimensionality. As discussed in \S~II.1, this quasi-two-dimensionalization outpaces the oscillatory forcing for Strouhal numbers $\mathit{Sr}<1/2$, while the conditions most favourable to transient growth and sustained turbulence occurred at $\mathit{Sr}$ order $10^{-2}$ to $10^{-3}$ (\S~\ref{sec:nlinm}). It is therefore likely that any initially 3D (large) scales would quasi-two-dimensionalize rapidly at the conditions of interest, and so decay each cycle as suggested by the present Q2D results, and as shown by the reduction in turbulent kinetic energy across all scales in \fig~\ref{fig:FoDNS}. Thus, although the impact of three-dimensionality in the smaller scales cannot be clarified without full 3D simulations, the ultimate issue is still one of sustaining turbulence at all scales, both Q2D and 3D.}

Finally, note that the subcritical turbulence observed for a steady base flow was far from fully developed \citep{Camobreco2021weakly}, and also exhibited temporal intermittency. Thus, it is still possible that fully developed turbulence may be reached at similar Reynolds numbers for both steady and pulsatile base flows. However, once at such high Reynolds numbers, there would appear to be no benefits, but perhaps also no drawbacks, to pulsatility. \textcolor{\mcy}{In addition, the Reynolds numbers required to sustain even intermittent turbulence for a steady base flow are already near-critical ($r_c \approx 1$) at the moderate Hartmann friction parameter $H=10$ investigated in this work, with supercritical Reynolds numbers ($r_c > 1$) required at larger $H$ \citep{Camobreco2020transition}. Thus, at the larger $H$ relevant to the motivation of this work (the cooling of fusion reactor blankets), neither a steady nor pulsatile base flow is likely to provide a route to Q2D turbulence through solely hydrodynamic or magnetohydrodynamic origins, e.g.~excluding instabilities driven by thermal convection. Instead, thermally driven instabilities \citep{Zikanov2013natural,Zhang2014mixed,Zhang2015turbulent}, or turbulence promoters \citep{Hussam2012enhancing,Cassels2016heat}, should be further considered at fusion relevant parameters.}


\begin{acknowledgments}
C.J.C.\ received support from the Australian Government Research Training Program (RTP). This research was supported by the Australian Government via the Australian Research Council (Discovery Grant DP180102647), the National Computational Infrastructure (NCI) and Pawsey Supercomputing Centre (PSC), by Monash University via the MonARCH HPC cluster and by the Royal Society under the International Exchange Scheme between the UK and Australia (Grant E170034).
\end{acknowledgments}
%


%

\end{document}